\documentclass[aps,prf,onecolumn,superscriptaddress,longbibliography,notitlepage]{revtex4-1}
\usepackage{graphicx}
\usepackage{dcolumn}
\usepackage{bm}
\usepackage{amsmath}

\begin{document}

\title{Determining the Onset of Hydrodynamic Erosion in Turbulent Flow}

\author{J.C. Salevan}
\affiliation{Department of Mechanical Engineering and Materials Science, Yale University, New Haven, Connecticut 06520, USA}
\author{Abram H. Clark}
\affiliation{Department of Mechanical Engineering and Materials Science, Yale University, New Haven, Connecticut 06520, USA}
\author{Mark D. Shattuck}
\affiliation{Benjamin Levich Institute and Physics Department, The City College of the City University of New York, New York, New York 10031, USA}
\affiliation{Department of Mechanical Engineering and Materials Science, Yale University, New Haven, Connecticut 06520, USA}
\author{Corey S. O'Hern}
\affiliation{Department of Mechanical Engineering and Materials Science, Yale University, New Haven, Connecticut 06520, USA}
\affiliation{Department of Physics, Yale University, New Haven, Connecticut 06520, USA}
\affiliation{Department of Applied Physics, Yale University, New Haven, Connecticut 06520, USA}
\author{Nicholas T. Ouellette}
\affiliation{Department of Civil and Environmental Engineering, Stanford University, Stanford, California 94305, USA}

\begin{abstract}
We revisit the longstanding question of the onset of sediment transport driven by a turbulent fluid flow via laboratory measurements. We use particle tracking velocimetry to quantify the fluid flow as well as the motion of individual grains. As we increase the flow speed above the transition to sediment transport, we observe that an increasing fraction of grains are transported downstream, although the average downstream velocity of the transported grains remains roughly constant. However, we find that the fraction of mobilized grains does not vanish sharply at a critical flow rate. Additionally, the distribution of the fluctuating velocities of non-transported grains becomes broader with heavier tails, meaning that unambiguously separating mobile and static grains is not possible. As an alternative approach, we quantify the statistics of grain velocities by using a mixture model consisting of two forms for the grain velocities: a decaying-exponential tail, which represents grains transported downstream, and a peaked distribution centered at zero velocity, which represents grains that fluctuate due to the turbulent flow but remain in place. Our results suggest that more sophisticated statistical measures may be required to quantify grain motion near the onset of sediment transport, particularly in the presence of turbulence.
\end{abstract}

\maketitle

\section{Introduction}
\label{sec:intro}
The transport of sediment by flowing water is a fundamental physical process with broad applications in fields ranging from geomorphology to land use and  agriculture~\cite{knisel1980,renard1997,jerolmack2010}. In many of these applications, the source of sediment is a bed of granular or other erodible material over which the water flows. If the stress delivered to the bed by the water is too weak, the bed will not be mobilized and it will be stable; but once the stress becomes large enough, the bed will become unstable and sediment will be entrained into the flow. Thus, understanding the conditions under which bed grains first start to move, often called incipient motion~\cite{buffington1997}, is of immense practical importance. Incipient motion is typically considered to be determined by the Shields number $\Theta$~\cite{shields1936}, which compares the shear stress $\tau$ exerted by the fluid on the bed to the buoyancy-reduced gravitational stress $(\rho_g - \rho_f) g d$, where $\rho_g$ and $\rho_f$ are the mass densities of the grains and fluid, respectively, $g$ is the gravitational acceleration, and $d$ is the diameter of a typical grain. The threshold for incipient motion is often quantified by measuring the sediment flux $q$~\cite{ouriemi2007,lajeunesse2010} as a function of $\Theta$ and identifying a critical value $\Theta_c$ above which grain velocities are nonzero.

However, this simple picture is incomplete for a variety of reasons. Both fluid forces~\cite{wiberg1987,kudrolli2016} and grain dynamics~\cite{clark2015hydro,clark2017} change qualitatively with the strength of the fluid driving, as quantified by the shear Reynolds number ${\rm Re}_*$ at the grain scale, meaning that there may not be a single, sharp transition to incipient motion~\cite{Paphitis2001}.  Previous studies have observed granular creep as the bed begins to mobilize~\cite{houssais2015}, suggesting a non-sharp transition, but also long transients~\cite{lobkovsky2008,hong2015}, suggesting an underlying bifurcation of some kind. Moreover, incipient motion is inherently stochastic due to disorder and variation in the bed structure~\cite{kirchner1990}, and turbulent fluctuations from the fluid begin to play a role at moderate to high Reynolds number~\cite{heathershaw1985,nezu1994,duran2012,yalin2015}. The presence of these fluctuations suggests that one may need to consider not only the mean stress but also the impulse imparted to grains via turbulent fluctuations~\cite{diplas2008,lee2012}. Thus, describing incipient grain motion involves mapping out the evolution of grain motion in a high-dimensional parameter space, and therefore is an ongoing scientific challenge. 

Here, based on the results of laboratory experiments in which we measured grain trajectories near incipient motion in a turbulent flow, we argue that since incipient motion is a stochastic process, it is best captured by a careful characterization of its statistics. We used high-speed imaging and particle tracking velocimetry to measure both the fluid flow and the grain motion. We quantify grain motion via the probability density function $P(u_g)$ of streamwise grain velocities $u_g$. We find that $P(u_g)$ is well described by a mixture model that is the sum of a function that is peaked at $u_g=0$ (which we fit to a Student's $t$-distribution), corresponding to grains that are not transported downstream but still fluctuate in place due to the turbulent flow, and a function with an exponential tail for positive $u_g$, corresponding to transported grains. We find that as $\Theta$ is increased above the threshold for incipient motion, the increase in sediment flux arises almost exclusively from an increased number of transported grains, while the average downstream velocity of mobile grains remains constant. Additionally, we find that the velocity distribution of non-transported grains becomes broader near the incipient motion threshold and thus the sediment flux does not vanish sharply. 

We begin below in Sec.~\ref{sec:methods} by describing our experimental methods, including the flow apparatus, the bed preparation protocols, and the measurement techniques. In Sec.~\ref{sec:results}, we describe our results, first comparing with previous studies that characterized incipient motion and then presenting our statistical analyses. Finally, in Sec.~\ref{sec:discussion}, we summarize and contextualize our results.

\section{Experiment}\label{sec:methods}

\subsection{Apparatus}

\begin{figure}
\centering
\includegraphics[width=.75\columnwidth]{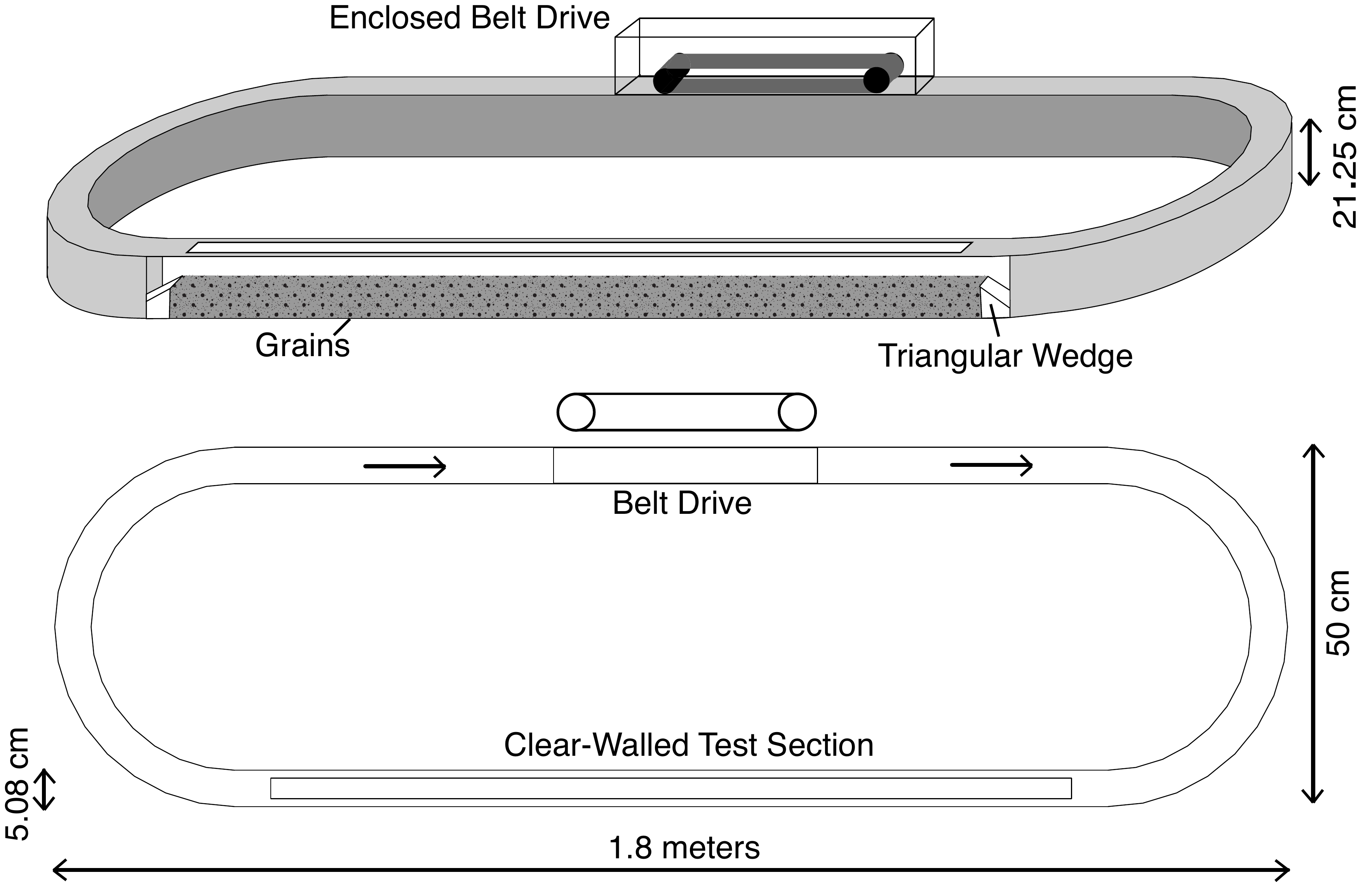}
\caption{Schematic of the experimental apparatus. The top panel shows a three-dimensional view, including the clear-walled test section with the grains and triangular wedges, and the enclosed, motorized belt drive on the opposite side. The bottom panel shows an overhead view, indicating the clockwise fluid flow direction. Overall, the apparatus is 1.8~m long, 21.25~cm tall, and 50~cm wide, with a 5.08~cm channel width. In the 1.2~m test section, the grain bed height is approximately 5.7~cm.}
\label{fig:apparatus}
\end{figure}

Our experimental apparatus is a closed, racetrack-shaped channel consisting of two straight sections joined by two U-shaped curves, as shown in Fig.~\ref{fig:apparatus}. The channel width is a uniform 5.08~cm, and its height is 21.25~cm. We drive the fluid flow with a toothed belt connected to a external rotary motor. The belt is fully immersed in water and is mounted to the top of one of the straight sections. The fluid velocity is controlled via the voltage sent to the motor controller, which produces a corresponding drive belt speed measured by an encoder on the belt roller axle. Additionally, we measure the fluid velocity in the test section directly (described in Sec.~\ref{sec:fluidflow}) and correlate that measurement with the belt speed. The test section is located on the opposite straight section, and is where data is collected on the grain motion and fluid flow. The entire apparatus is fully enclosed, so that there is no free surface. The side walls of the test section are made of tempered glass and the top surface of the test section is made of clear acrylic, allowing optical access for imaging.

To prepare our experiments, we first add soda lime glass beads with diameter $d = 0.1125 \pm 0.0125$~cm and density $\rho_g = 2.5$~g/cm$^{3}$ to the test section, up to a fill height of roughly 5.7~cm (about 51 grain diameters). This erodible bed is constrained in the streamwise direction by triangular wedges, which both help to slow the formation of large-scale bedforms as the experiments proceed and keep most of the grains in the test section. After the grains are added, we completely fill the apparatus with water. We then drive the fluid at a slow speed (far below the threshold for grain motion) to dispel trapped air bubbles in the system. This weak fluid forcing also allows the beads on the surface of the bed to settle slightly, creating a more uniform initial bed state.

After the initial state is prepared, we quasistatically increase the free stream fluid velocity in increments of approximately 0.95 cm/s up to a maximum velocity for a given experimental run. After each increment, we allow the flow to stabilize, and then acquire data. Once we reach our maximum velocity, we slowly decrease the fluid velocity in the same manner, taking more data after each step. We reset the initial bed state as described above before each set of experiments. The granular bed and the fluid flow are imaged separately: one camera views the bed from overhead, through the top of the test section, to image the bed surface, and a second camera views the fluid flow through a side window, to image the velocity profile of the fluid flow. 

\subsection{Fluid Flow and Shear Stress}
\label{sec:fluidflow}

To quantify the fluid flow, we add fluorescent red polyethylene microspheres that are 75-90 $\mu$m in diameter and density-matched with water. We illuminate a vertical slice in the center of the test section with a light sheet produced by a 532~nm laser and a cylindrical lens mounted above the apparatus. Particles in the light sheet fluoresce, and are imaged using a high-speed camera at a frame rate of 1400 frames per second and a spatial resolution of 181 pixels/cm. We record 5-second videos through the side of the test section, and track the tracer particles via a multiframe predictive particle-tracking algorithm \cite{ouellette2006a}. Accurate velocities are then computed from the particle trajectories via convolution with a smoothing and differentiating kernel \cite{mordant2004}.

\begin{figure}
\centering
\includegraphics[width=.7\columnwidth]{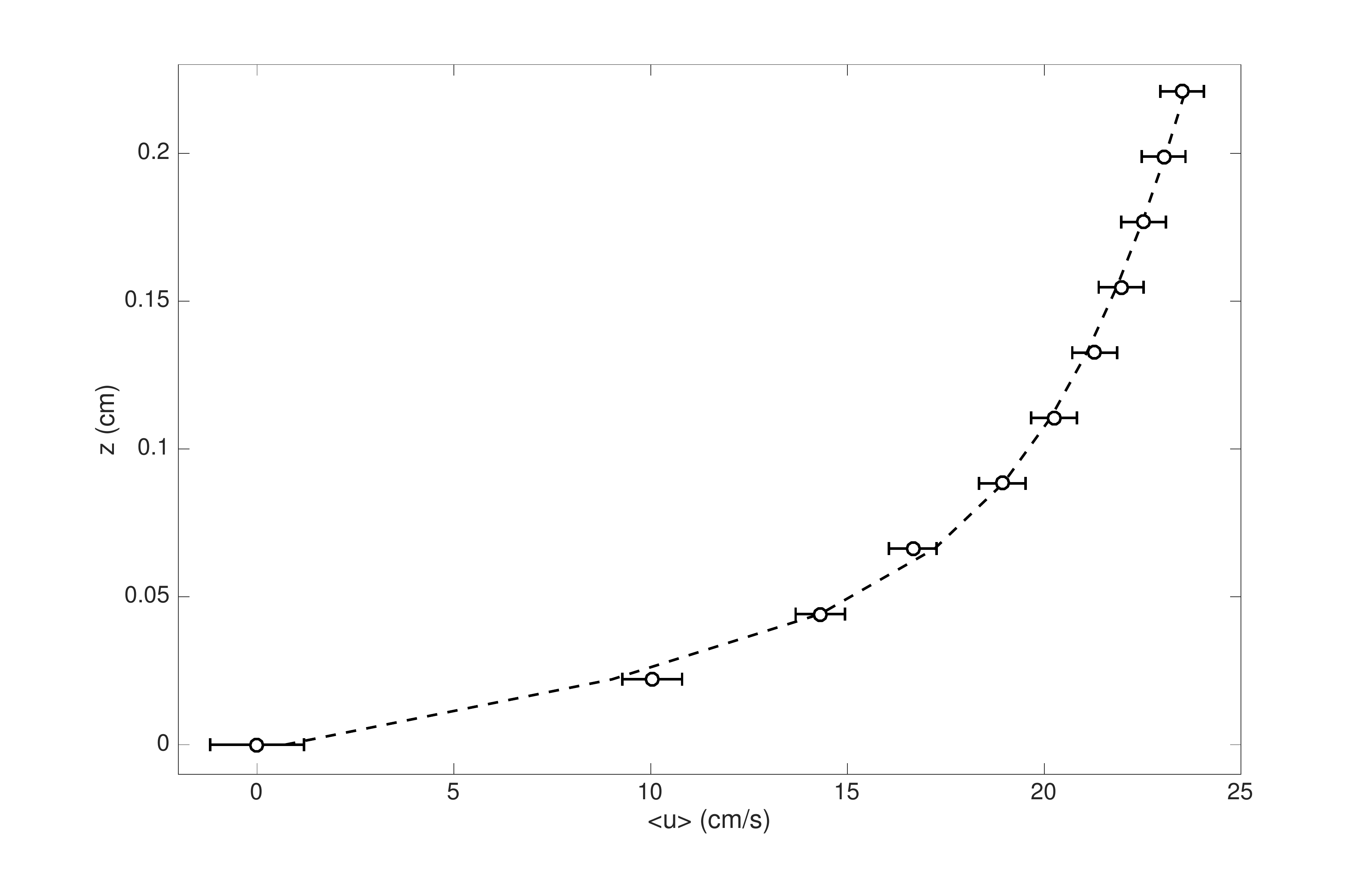}
\caption{A typical fluid velocity profile, as measured from the streamwise velocities $u$ of tracer particles in the fluid. Each data point (open circles) represents a time average over all tracer particle trajectories for binned values of the height. The error bars show the standard deviation of the velocities in each bin. The dashed line is a fit following the procedure described by Rodr\'iguez-L\'opez \textit{et al.}~\cite{rodriguez2015}, from which we extract the wall shear stress $\tau_{wall}$ (see text).}
\label{fig:profile}
\end{figure}

In order to extract the hydrodynamic stress exerted on the granular bed, we first construct the mean vertical velocity profile by binning the measured velocities in height (with at least 20,000 samples per bin). An example profile is shown in Fig.~\ref{fig:profile}, where the error bars show the standard deviation of the velocities in each bin. To compute the wall shear stress $\tau_{wall}$, we fit the velocity profile using the modified logarithmic law method described by Rodr\'iguez-L\'opez \textit{et al.}~\cite{rodriguez2015}, the result of which is shown as the dashed curve in Fig.~\ref{fig:profile}. The wall shear stress is then given as $\tau_{wall} = \mu\frac{\partial u}{\partial y}$ evaluated at $y=0$, where $\mu$ is the dynamic viscosity of water. $\tau_{wall}$ in turn allows us to compute the friction Reynolds number ${\rm Re}_\tau = \delta / \delta_\nu$, where $\delta$ is the channel half-height, $\delta_\nu = \nu \sqrt{\rho_f / \tau_{wall}}$ is the viscous length scale \cite{pope2000}, and $\nu = \mu / \rho_f$ is the kinetic viscosity of water. For the results presented here, $450 < {\rm Re}_\tau < 2200$. 

The wall shear stress also allows us to compute the Shields number $\Theta$ as 
\begin{equation} 
\Theta = \frac{\tau_{wall}}{(\rho_g - \rho_f)gd}, \\
\label{eq:shields}
\end{equation}
where in our experiments the Shields numbers are in the range $0.005 < \Theta < 0.045$. The shear Reynolds number at the grain scale ${\rm Re}_{*} = d/\delta_\nu$ is also used to specify the sediment-transport regime, and in our experiments, we have $6 < {\rm Re}_* < 30$. 

\subsection{Granular Bed}
\label{sec:grainflow}

\begin{figure}
\centering
\includegraphics[width=.9\columnwidth]{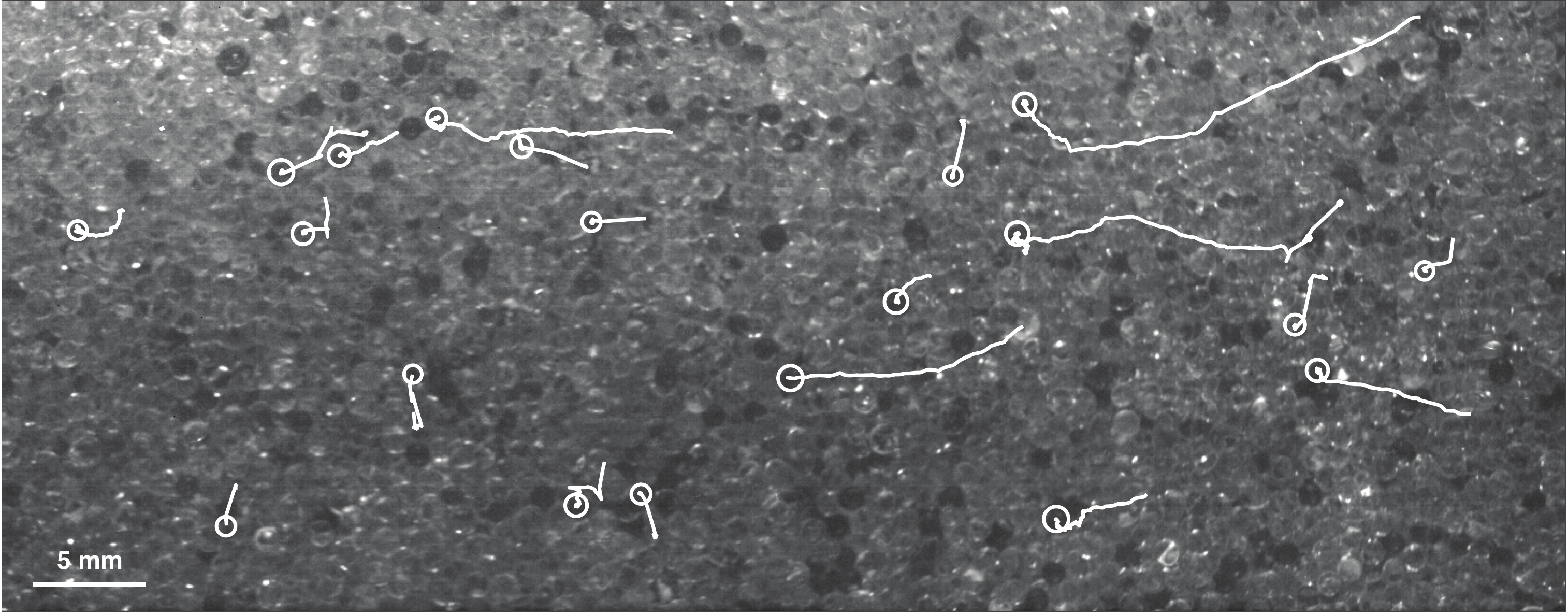}
\caption{A raw image of the granular bed shown from above, with selected grain trajectories overlaid in white. This snapshot shows white circles to indicate the position and radius of a selected subset of tracked grains during a single frame, with trailing lines behind each grain indicating the motion from the beginning of each track. The time interval for each track varies, as the time at which the Lagrangian particle tracking algorithm began to track each specific grain also varies. We include a 5~mm scale bar on the bottom left, and the fluid flows from right to left.}
\label{fig:overhead}
\end{figure}

To quantify the motion of bed grains, about 5\% of the glass beads in the bed are colored black, which appear dark against the bright background of illuminated, clear beads. Aside from the color, the black beads are identical to the clear beads. We illuminate the test section volume with an array of LEDs, and use a second camera positioned above the bed to take video of the bed surface at 250-500 frames per second with a resolution of 190 pixels/cm. This frame rate and resolution ensure that beads move no more than half of a grain diameter per frame over the range of driving flows we consider.

After normalizing each image to correct for lighting inhomogeneities, we use a circular Hough transform to locate the position of black beads on the bed surface accurately. With those positions, we then use the same Lagrangian particle tracking algorithm as used for the fluid flow to extract trajectories for each bead. 
Fig.~\ref{fig:overhead} shows an example raw image frame from a video of the granular bed. Selected grain trajectories are shown in white, with the current position of each grain in the frame shown as a white circle, indicating both the position and the radius extracted from the Hough transform.

\section{Results}
\label{sec:results}

\begin{figure}
\raggedright
(a) \hspace{78mm} (b) \\
	\includegraphics[width=0.49\textwidth]{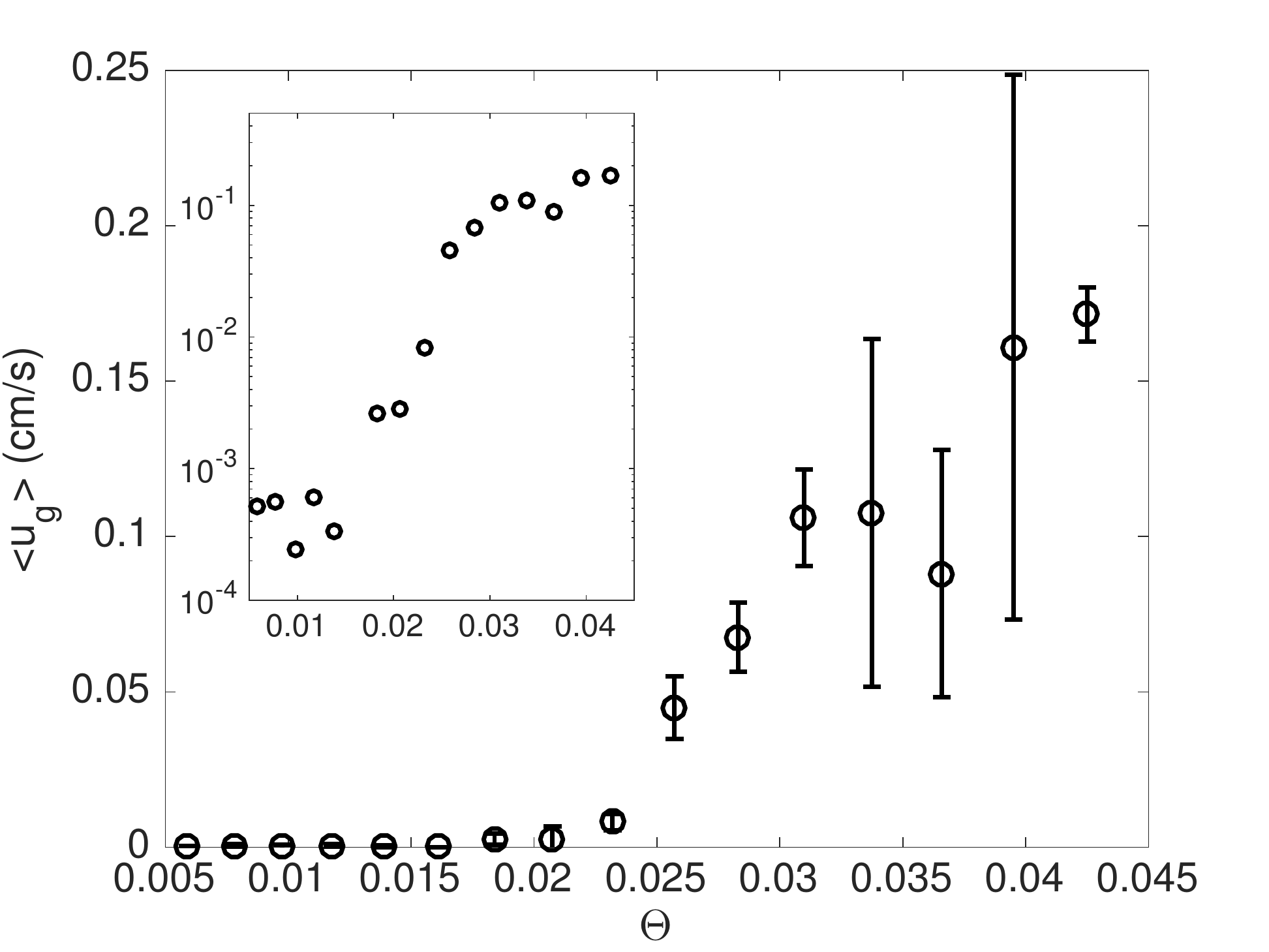}
~
\includegraphics[width=0.49\textwidth]{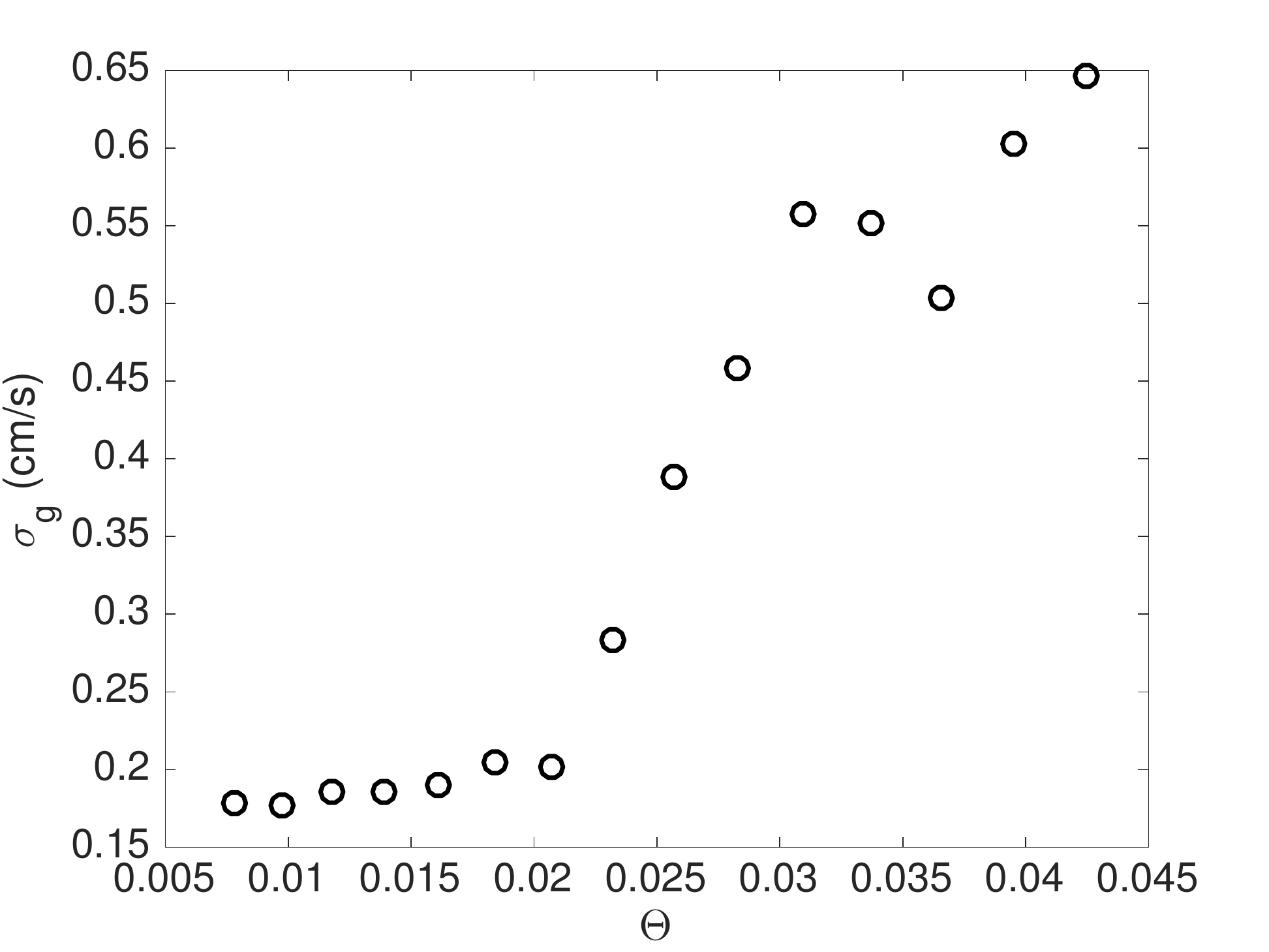}
\caption{(a) The mean grain velocity $\langle u_g \rangle$ as a function of Shields number $\Theta$. The error bars show the standard error computed for six trials, and the inset shows the same data plotted on semilogarithmic axes. (b) The standard deviation of grain velocity $\sigma_g$ as a function of $\Theta$.}
\label{fig:meanvel}
\end{figure}

Incipient motion is typically defined as the point at which the sediment flux $q$ vanishes. If all grain motion is associated with downstream transport, then the mean grain velocity $\langle u_g \rangle$ should be proportional to $q$; thus, since $\langle u_g \rangle$ is typically easier to measure, it is often used as a proxy for $q$ in determining the onset of sediment transport. In Figure~\ref{fig:meanvel}(a), we plot $\langle u_g \rangle$ as a function of $\Theta$, where in our case the average is taken over all measured velocities for each tracked grain. For small $\Theta$, $\langle u_g \rangle$ is approximately zero. For $\Theta > 0.023$, we observe that $\langle u_g \rangle$ grows quasi-linearly. For $0.018 < \Theta < 0.023$, $\langle u_g \rangle$ smoothly varies between these two limits. In addition to the mean grain velocity, we can also investigate the standard deviation of grain velocity $\sigma_g$, since we would expect variation among the grains and in time. Figure~\ref{fig:meanvel}(b) shows $\sigma_g$ for the grain velocity as a function of $\Theta$. Although it varies in a qualitatively similar way as $\langle u_g \rangle$ with $\Theta$, $\sigma_g$ does not vanish for small $\Theta$; instead, the magnitude of the fluctuations in grain velocity remain finite. This result suggests that even below the onset of net sediment transport, grains can rattle on the surface of the bed.

Measuring the mean grain velocity is not the only method that has been used to determine the onset of sediment transport. For example, several recent studies~\cite{lajeunesse2010,roseberry2012,gonzalez2016} have taken the approach of first choosing a velocity threshold to identify which grains are in motion. The ratio of the number of these ``active'' grains to the total number of grains or the bed surface area is known as the particle or grain activity. The critical Shields number $\Theta_c$ can ideally be determined by identifying the point at which the grain activity vanishes~\cite{lajeunesse2010}. After identifying mobilized grains, one can examine the statistics of the velocities of moving grains, since the grain flux $q$ is proportional to the product of the grain activity and the mean velocity. The probability density function (PDF) $P_m(u_g)$ of the velocities $u_g$ of mobile grains is often found to be close to exponential~\cite{lajeunesse2010,roseberry2012,furbish2013,fathel2015}, so that it is well fit by
\begin{equation}
P_m(u_g) = \frac{1}{u_g^*} e^{-u_g / u_g^*}.
\label{eq:exp}
\end{equation}
Here, $u_g^*$ is the average velocity of the mobile grains. This distribution can be derived from maximum-entropy statistical-mechanical arguments, starting from a microcanonical ensemble and constraining the total grain momentum~\cite{furbish2013}. Equation~\eqref{eq:exp} is particularly successful for describing bedload transport when the number of moving grains becomes large~\cite{furbish2013,fathel2015}. However, this entire approach is predicated on first identifying moving grains. But very close to the threshold of incipient motion, the grain activity and subsequent velocity statistics may become highly sensitive to the choice of threshold, especially when the driving flow is turbulent. In this case, the transition to sustained sediment transport may not be sharp, but may be blurred by the turbulent fluctuations in the driving flow.

\begin{figure}[t]
\raggedright
 \hspace{0mm}(a) \hspace{78mm} (b) \\
	\centering
	\includegraphics[width=0.49\textwidth]{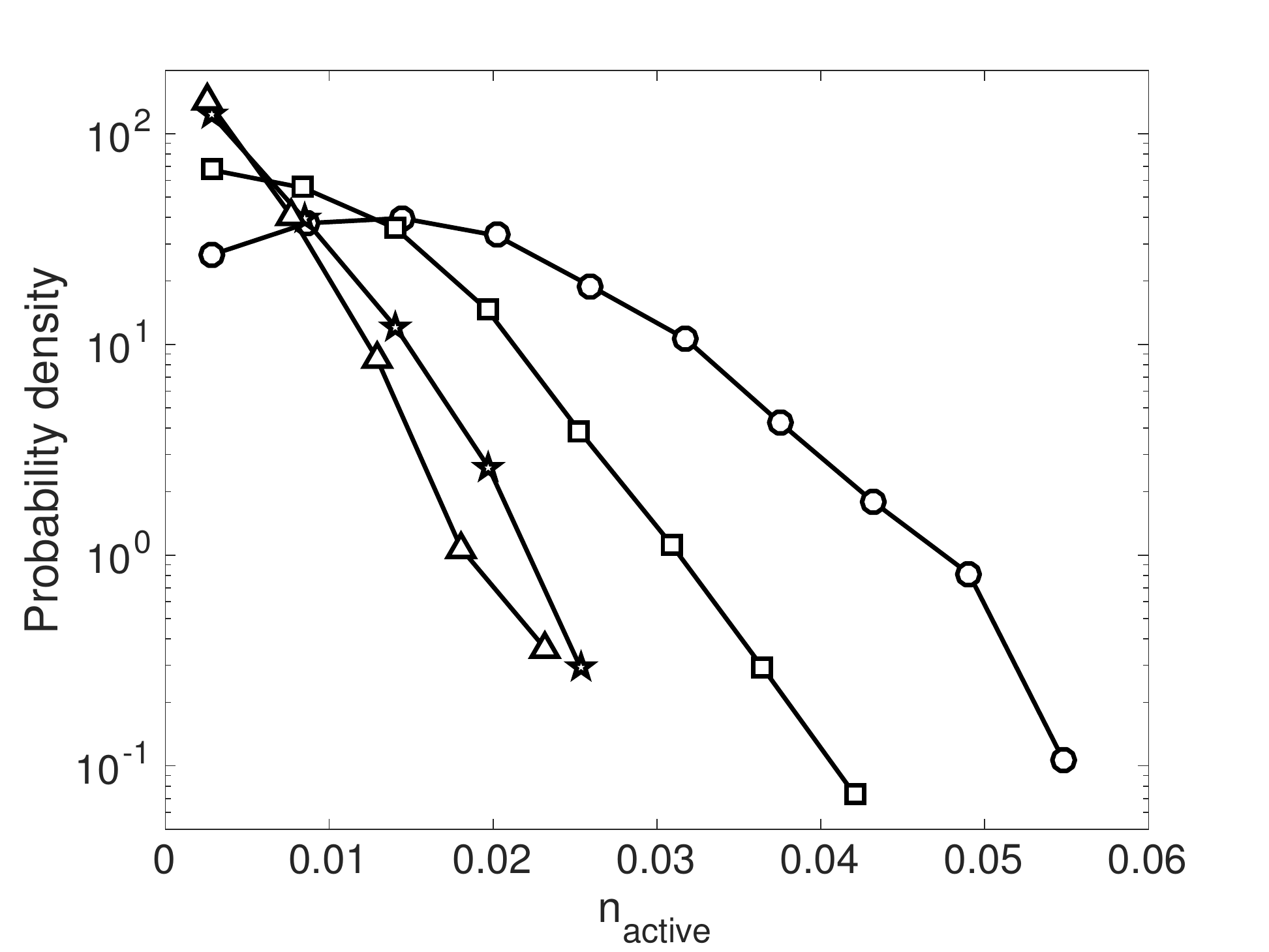}
	\includegraphics[width=0.49\textwidth]{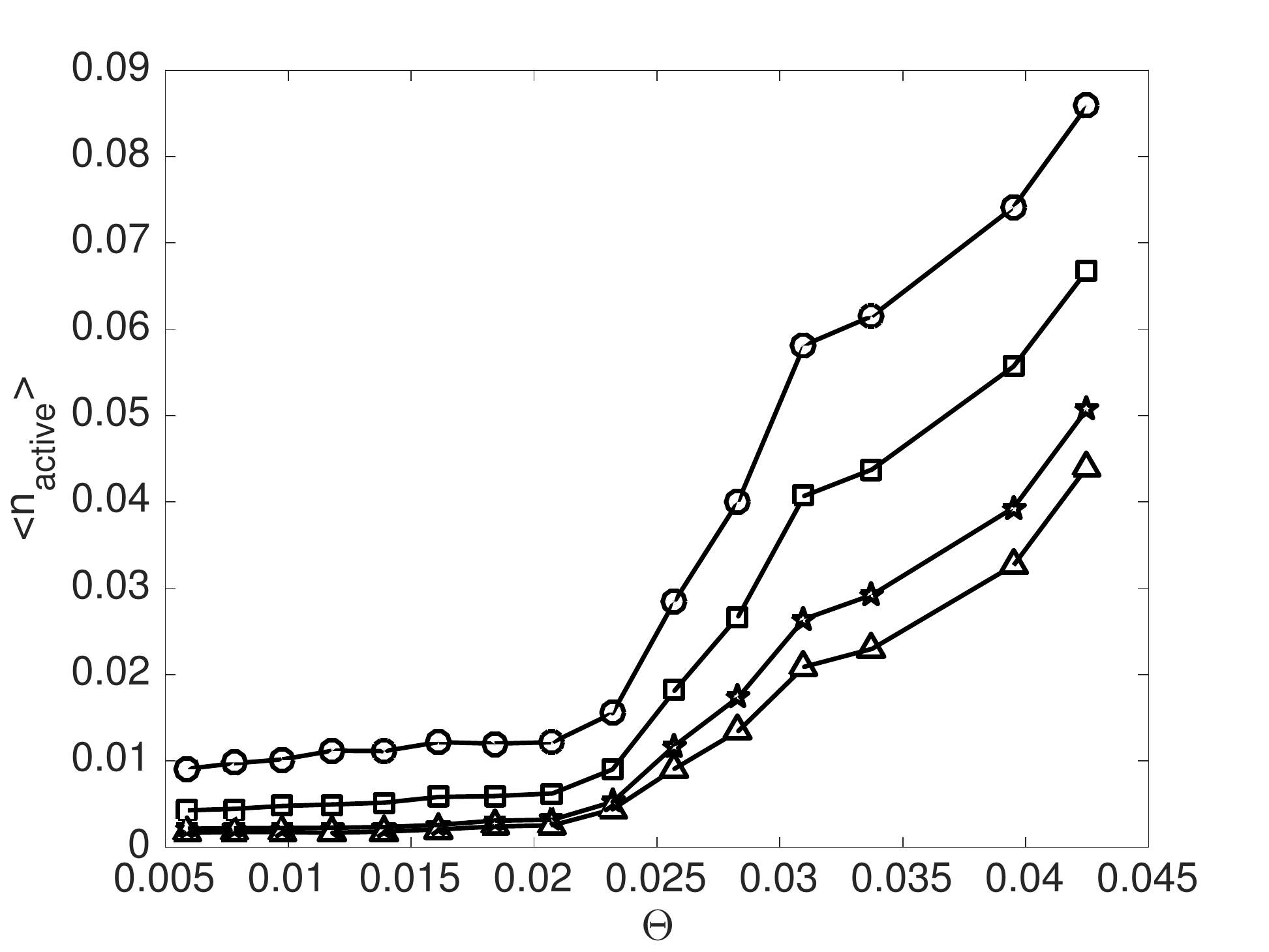}
\caption{(a) The probability density that a fraction $n_{\rm active}$ of all tracked grains are moving at any given time for $\Theta = 0.023$, plotted for four different values of the velocity threshold $u_{\rm threshold}$ defining grain motion: 0.75~cm/s (circles), 1~cm/s (squares), 1.5~cm/s (stars), and 2~cm/s (triangles). These PDFs are highly sensitive to the choice of threshold. (b) The mean fraction of active grains for the same values of $u_{\rm threshold}$ as a function of $\Theta$.}
	\label{fig:activityonset}

\end{figure}

Figure~\ref{fig:activityonset}(a) shows activity distributions from our experiments. Each curve shows a PDF of the fraction of active grains $n_{\rm active}$ in any given video frame from all experiments at a given value of $\Theta$, where active grains are defined as those that have an instantaneous streamwise velocity greater than a threshold velocity $u_{\rm threshold}$. These PDFs are sensitive to the choice of $u_{\rm threshold}$. We plot the mean fraction of active grains $\langle n_{\rm active} \rangle$ as a function of $\Theta$ in Fig.~\ref{fig:activityonset}(b). The behavior is qualitatively similar to the mean velocity shown in Fig.~\ref{fig:meanvel}(a), but $\langle n_{\rm active} \rangle$ remains non-zero in general for $\Theta<\Theta_c$. The value of this subcritical plateau (for $\Theta<\Theta_c$) is dependent on $u_{\rm threshold}$.

\begin{figure}
\centering
\includegraphics[width=0.7\textwidth]{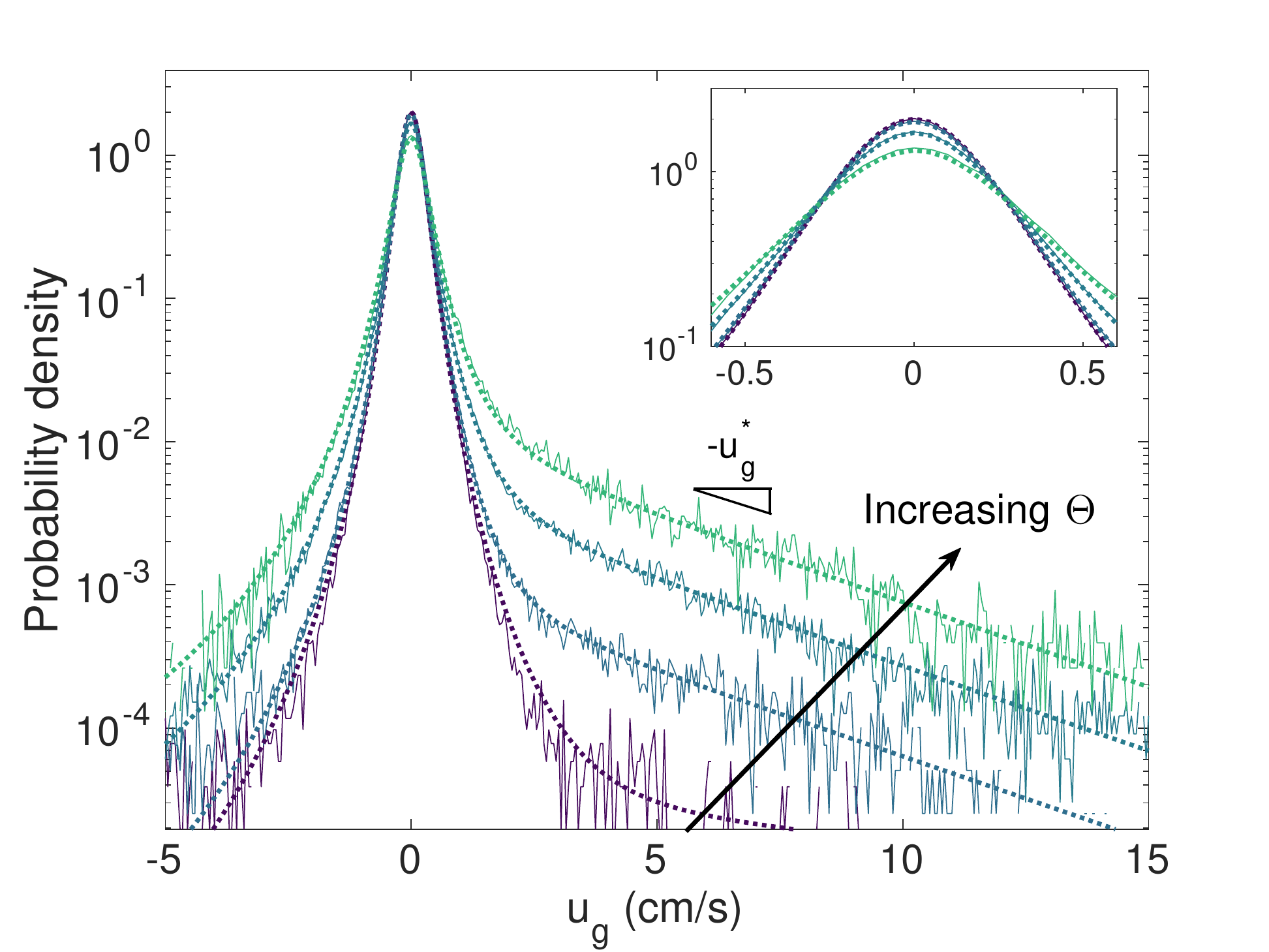}
\caption{\raggedright PDFs $P(u_g)$ of grain velocities $u_g$ (solid curves), for $\Theta = 0.0059$, 0.0232, 0.0283, and 0.0425. Dashed curves show fits of the data to Eq.~\eqref{eq:vel-pdf}. The inset shows a close-up of the same data near the core of the PDF.}
\label{fig:pdf}
\end{figure}

It is not surprising that mobile and non-mobile grains are difficult to separate, especially very near the onset of sustained sediment transport. This blurring arises from physical effects, primarily turbulence, as well as some contribution from finite experimental measurement accuracy. Thus, instead of identifying only mobile grains and examining their statistics, we examine \emph{all} tracked grains together. In Fig.~\ref{fig:pdf}, we show PDFs of all tracked grain velocities for several values of $\Theta$. The solid lines show measured data, and the dashed lines are a model that is fit to the data, as described below. Several features of these PDFs are noteworthy. As one might expect, most of the measured particle velocities are clustered near zero. These points are from bed particles that we identify but that are not mobile, since near incipient motion most of the bed is still not moving. However, the peak at $u_g = 0$ is not sharp but rather is somewhat broad, even in the low-$\Theta$ limit, which arises from a combination of physical fluctuations and experimental noise. At larger values of $u_g$, the PDF becomes a decaying exponential with a characteristic velocity scale $u_g^*$, as in Eq.~\ref{eq:exp}. This part of the PDF arises from mobilized grains. Between the static and mobile parts of the PDF, there is a crossover region that does not fit well either with an exponential or with a normal distribution centered at $u_g = 0$. In addition, the magnitude of $P(u_g)$ for small velocities increases with $\Theta$ for both positive and negative $u_g$. Visual inspection of the crossover region for $u_g>0$ suggests that $u_{\rm threshold}$ should be chosen between 1 and 4~cm/s, but there is no objective way to choose within this range. Moreover, the width of the crossover region appears to grow with $\Theta$, meaning that the appropriate choice for $u_{\rm threshold}$ would vary with $\Theta$.

These PDFs clearly show why activity and mean velocity are difficult to use as precise indicators of the onset of net sediment transport: the development of the exponential tail that characterizes downstream flux is gradual, and occurs at the same time as the central peak broadens due to turbulence and noise. As an alternative way to quantify our data over the full range of $\Theta$, we instead here suggest a fit of the \emph{entire} PDF via a mixture model. As we have already indicated, the tail at positive $u_g$ ought to be exponential. The central core due to turbulence is quasi-Gaussian; we, however, find better agreement (and are better able to fit the entire PDF) if we use a Student's $t$-distribution instead. Thus, we fit the grain velocity PDFs to a function of the form
\begin{equation}
\label{eq:vel-pdf}
P(u_g) = A \frac{\Gamma\left(\frac{\zeta + 1}{2}\right)}{\sigma \sqrt{\zeta \pi} \Gamma\left(\frac{\zeta}{2}\right)} \left[ \frac{\zeta + \left(\frac{u_g}{\sigma}\right)^2}{\zeta} \right]^{\left(- \frac{\zeta+1}{2}\right)} + B \frac{1}{u_g^*} e^{-u_g / u_g^*}.
\end{equation}
Here, $A$ and $B$ are mixture fractions that sum to unity (to keep the overall PDF normalized) and give the relative fraction of particles in the (static) core and the (mobile) tail; $\Gamma$ is the gamma function; $\sigma$ is the characteristic width of the $t$-distribution and $\zeta$ is its shape parameter that sets the heaviness of its tails. When $\zeta$ is small, the tails are heavier than a Gaussian with the same variance, and as $\zeta\rightarrow\infty$, the $t$-distribution becomes a Gaussian with standard deviation $\sigma$.

\begin{figure}
\raggedright
 (a) \hspace{78mm} (b) \\
	\centering
\includegraphics[width=0.49\textwidth]{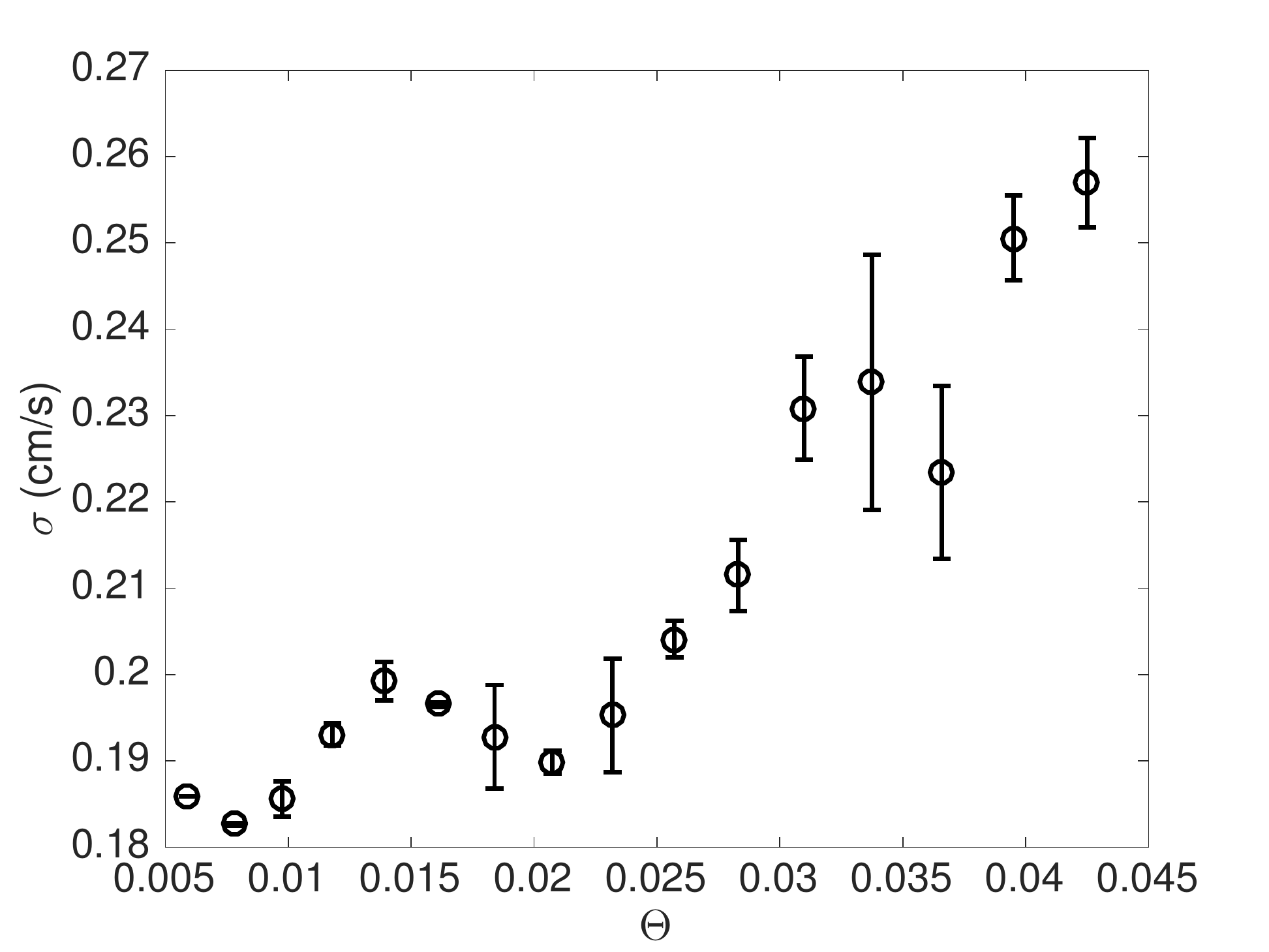}
~
\includegraphics[width=0.49\textwidth]{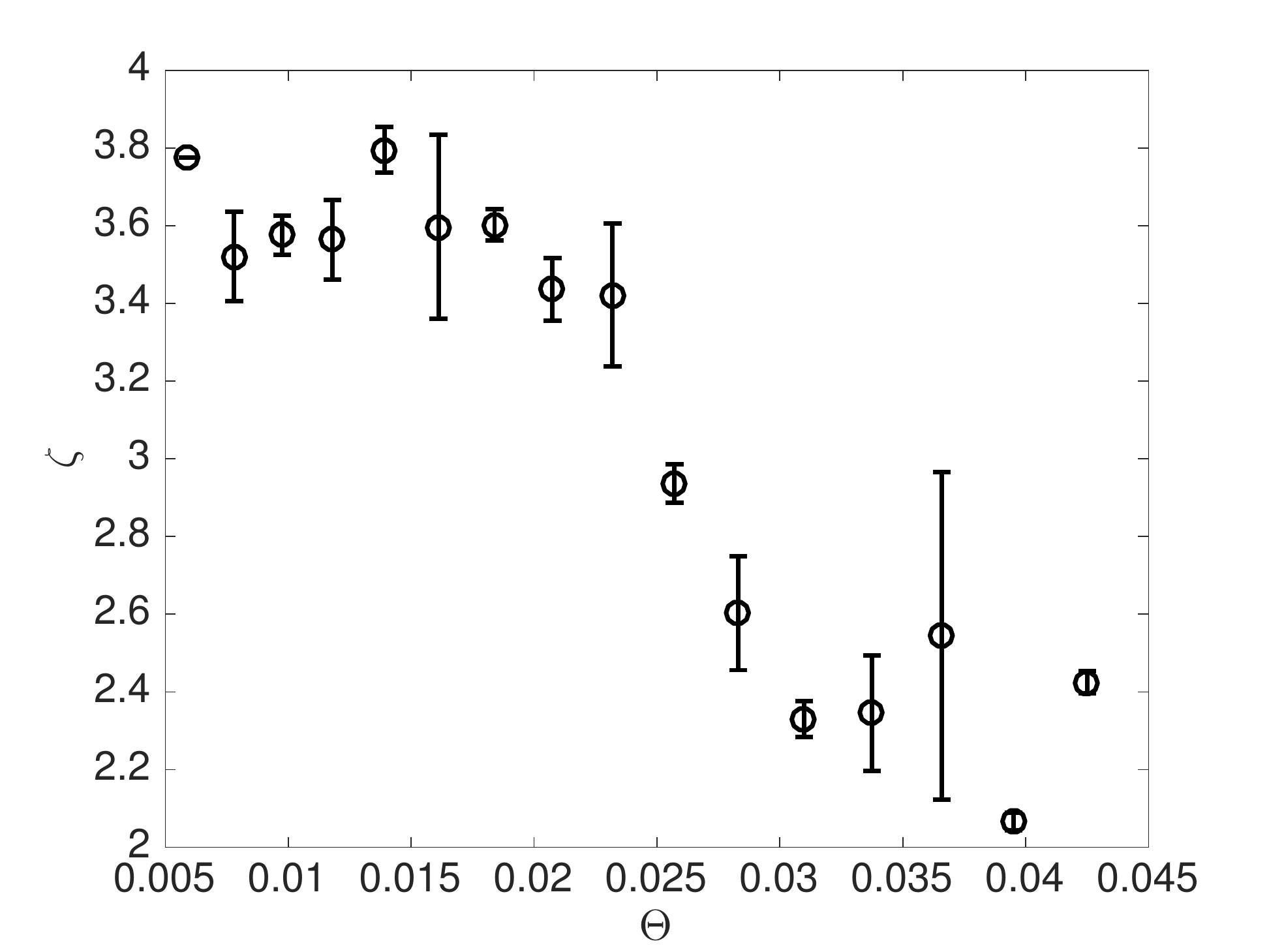} \\
\raggedright
 (c) \hspace{78mm} (d) \\
	\centering
\includegraphics[width=0.49\textwidth]{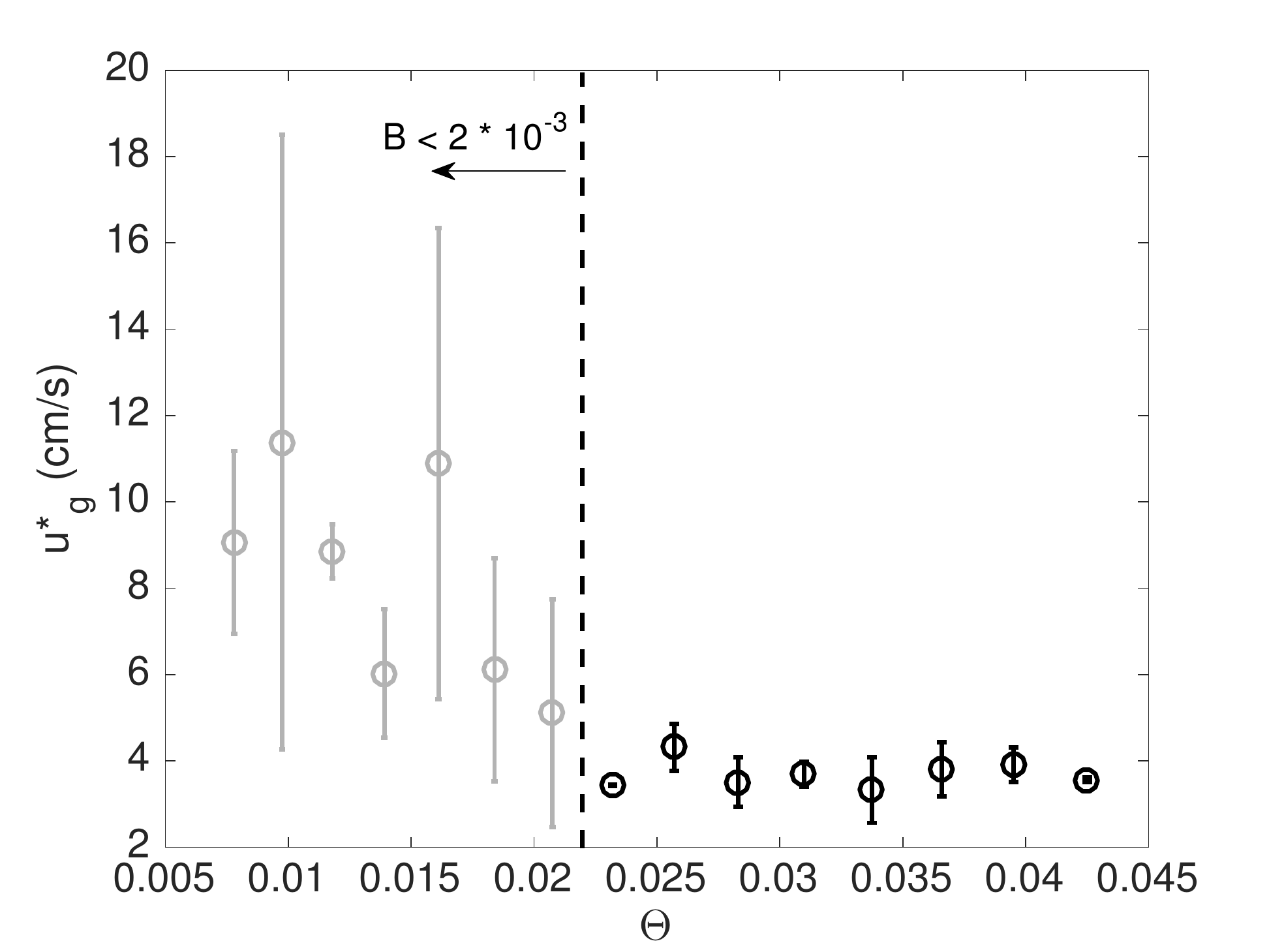}
\includegraphics[width=0.49\textwidth]{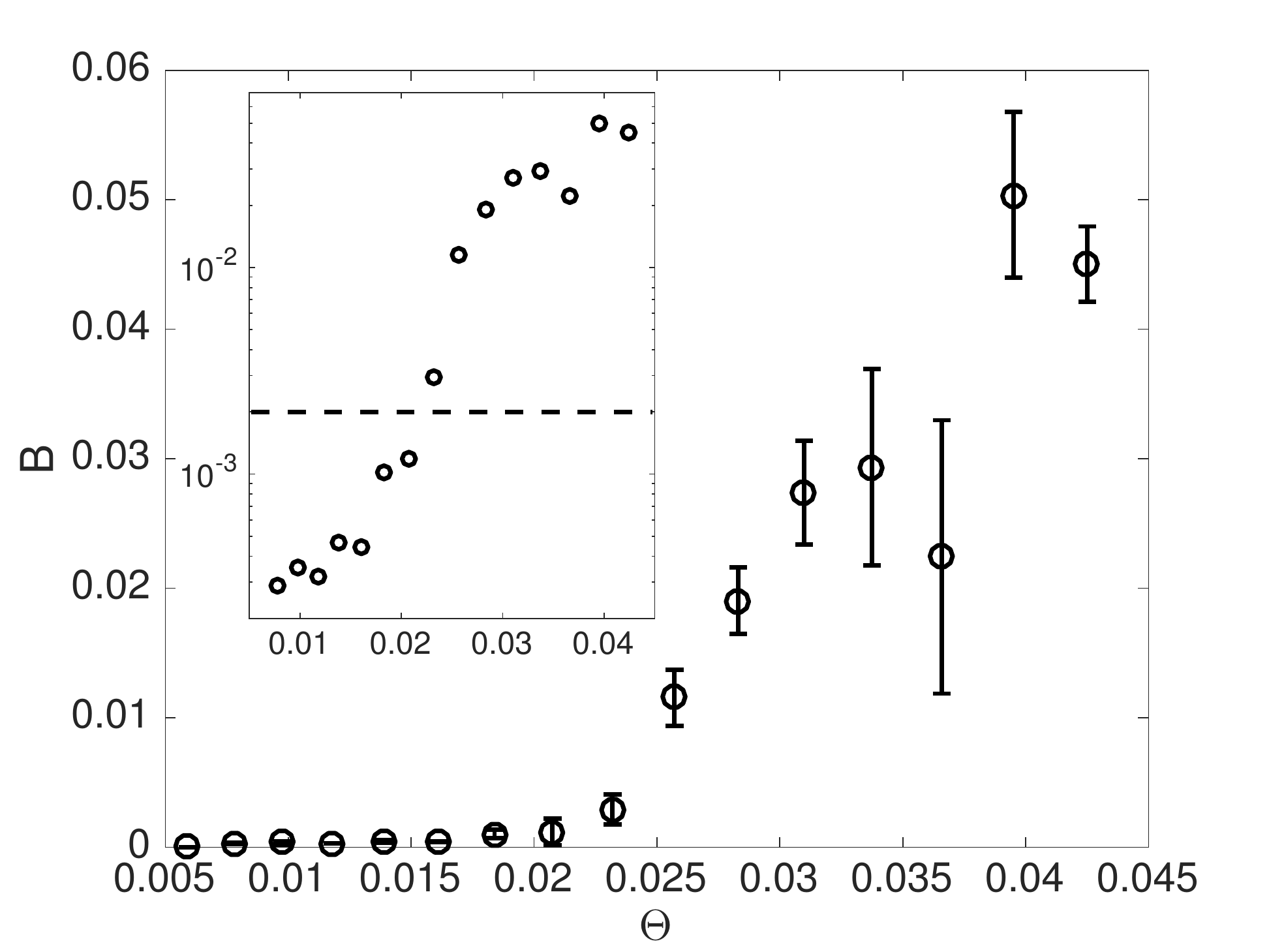}
~
\\ \raggedright
\caption{\raggedright Fit parameters (a) $\sigma$, (b) $\zeta$, (c) $u_g^*$, and (d) $B$ from Eq.~\eqref{eq:vel-pdf} as a function of Shields number $\Theta$. In all cases, the error bars are the standard error computed over six different experimental trials. We mark on panel (c) points (in gray) at which $B<2\times10^{-3}$, where our fit for $u_g^*$ is no longer physically meaningful. The inset to panel (d) shows the same data plotted on semi-logarithmic axes, with a horizontal line at $B=2\times10^{-3}$.}
\label{fig:fit-params}
\end{figure}

Figure~\ref{fig:fit-params} shows the parameters $\sigma$, $\zeta$, $u_g^*$, and $B$ that characterize Eq.~\eqref{eq:vel-pdf} as a function of $\Theta$. Although we allow both $A$ and $B$ to vary in our fits, we always recover $A+B = 1\pm 0.04$. Figure~\ref{fig:fit-params}(a) shows the characteristic width $\sigma$ of the $t$-distribution. For $\Theta<\Theta_c$, $\sigma$ fluctuates somewhat but remains roughly constant. For $\Theta>\Theta_c$, $\sigma$ grows monotonically. We note that we observe identical behavior for $\sigma$ when we fit the core to a normal distribution, where $\sigma$ is simply the standard deviation. Additionally, as shown in Fig.~\ref{fig:fit-params}(b) we observe a decrease in $\zeta$ for $\Theta>\Theta_c$, meaning that the tails of the distribution are becoming heavier for both positive and negative $u_g$. We interpret this effect as a manifestation of the increasing turbulence (since in our experiment the Shields and Reynolds numbers vary together): the more intense turbulence implies stronger fluctuations in wall shear stress and velocity, which in turn lead to more rapid fluctuations of non-mobilized bed grains. We note that the $t$-distribution is symmetric about zero, so the agreement of our model with the data suggests that the turbulence-driven fluctuations of the grains have similar positive and negative downstream values. By definition, the grains described by the $t$-distribution do not contribute to the net transported flux.

For larger grain velocities, we find excellent agreement with a decaying exponential, as in Eqs.~\eqref{eq:exp} and \eqref{eq:vel-pdf}. Figure~\ref{fig:fit-params}(c) shows $u_g^*$, which can be interpreted as the mean velocity of the transported grains. For $\Theta > \Theta_c$, $u_g^*$ remains roughly constant at $u_g^* \approx 3.5$~cm/s. Thus, near the onset of net sediment transport, increases in the grain flux appear to be primarily driven not by increasing the velocity of mobilized grains but by increasing their number. This conclusion is supported by Fig.~\ref{fig:fit-params}(d), which shows that $B$, which can be roughly interpreted as the fraction of grains that fall into the exponential tail, grows roughly linearly for $\Theta > 0.023$. However, $B$ does not vanish sharply at a particular value of $\Theta$. This behavior can be more clearly seen in the inset of Fig.~\ref{fig:fit-params}(d), which follows the same trend of $\langle u_g \rangle$, shown in Fig.~\ref{fig:meanvel}(a). For $\Theta < 0.018$, $B$ vanishes to within our experimental accuracy. For $0.018 < \Theta < 0.023$, we observe a transition between $B\approx 0$ for $\Theta<0.018$ and linear growth with $\Theta$ for $\Theta>0.023$. Thus, the transition to net sediment transport is blurred, presumably by the turbulence in our experiment. We also note that the numerical values for $B$ in Fig.~\ref{fig:fit-params}(d) are similar to the values of $\langle n_{\rm active} \rangle$ plotted in Fig.~\ref{fig:activityonset}(b). However, fitting to Eq.~\eqref{eq:vel-pdf} allows us to separate transported grains from fluctuating grains, as $B$ vanishes for small $\Theta$ while the fraction of active grains does not. Note that as $B$ vanishes, our fit for $u_g^*$ is no longer physically meaningful, as there are no transported grains. Thus, although we include data for $u_g^*$ for all $\Theta$, we note in Fig.~\ref{fig:fit-params}(c) points at which $B<2\times10^-3$.

\section{Discussion and Conclusions}
\label{sec:discussion}

Here we have presented experimental measurements of the statistics of grain motion across the transition to net sediment transport in a laboratory flow. We quantify the grain motion and the fluid flow using high-speed video and particle tracking. We considered several methods for attempting to pinpoint the onset of sediment transport, and argued that fitting a mixture model to the full PDF of grain velocities gives the most objective way of accomplishing this goal, particularly in the presence of turbulence. Appealingly, this method does not require the choice of a threshold for deciding which grains are contributing to net sediment transport and which are simply rattling in the presence of turbulent fluctuations; instead, the different statistical properties of these two classes of grains are used directly to distinguish them.

Using this approach, we showed that the density of grains transported downstream does not vanish sharply at a well-defined critical Shields number $\Theta_c$. Instead, we find a crossover region between $0.018 < \Theta < 0.023$, which is similar to $\Theta_c$ as measured in Ref.~\cite{lajeunesse2010}. Thus, our results indicate that there may not always be a clear separation between mobile and static grains: in the PDFs of grain velocities, we find a crossover region between transported grains and immobile grains that changes as $\Theta$ is varied. Here, we attribute this ``blurring'' of the transition to sediment transport as an effect of the turbulent fluctuations; we note, however, that this kind of continuous rather than sharp transition was also recently reported in the laminar case and interpreted as resulting from granular creep \cite{houssais2015}. We also find that as $\Theta$ is increased above the onset of net downstream grain motion, the fraction of grains that are mobilized increases although the mean velocity of mobilized grains remains roughly constant. This result is consistent with the findings in Ref.~\cite{roseberry2012}, but different from those in Ref.~\cite{lajeunesse2010}, where both activity and mean velocity of mobile grains apparently increased with increasing $\Theta$, albeit at higher transport rates. Further work is therefore required to settle this question, and to understand whether there are additional physical effects that distinguish these situations.

\begin{acknowledgments}
This research was sponsored by the Army Research Laboratory and was
accomplished under Grant Numbers W911NF-14-1-0005 and W911NF-17-1-0164
(A.H.C., N.T.O., and C.S.O.). The views and conclusions contained in
this document are those of the authors and should not be interpreted
as representing the official policies, either expressed or implied, of
the Army Research Laboratory or the U.S. Government. The
U.S. Government is authorized to reproduce and distribute reprints for
Government purposes notwithstanding any copyright notation
herein. This work was also supported by the National Science Foundation (NSF) under the Graduate Research Fellowship Program (GRFP) Grant No.~ DGE-1122492 (J.C.S.) and by the National Science Foundation (NSF) Grant No.~CMMI-1463455 (M.D.S.). 
\end{acknowledgments}

\bibliography{erosion_sim3_nto}

\end{document}